\documentstyle[prb,aps,epsf,psfig]{revtex}
\begin{document}
\twocolumn[\hsize\textwidth\columnwidth\hsize
           \csname @twocolumnfalse\endcsname
\title{Bernoulli potential at a superconductor surface}
\author{Pavel Lipavsk\'y, Jan Kol{\'a}{\v c}ek and 
Ji{\v r}{\'\i}~J.~Mare{\v s}}
\address{Institute of Physics, Academy of Sciences, 
Cukrovarnick\'a 10, 16258 Praha 6, Czech Republic}
\author{Klaus Morawetz}
\address{Max-Planck-Institute for the Physics of Complex 
Systems, Noethnitzer Str. 38, 01187 Dresden, Germany}
\maketitle
\begin{abstract}
The electrostatic Bernoulli potential measured at the surface
of a superconductor via Kelvin capacitive coupling is shown to
be independent of the pairing mechanism. This contrasts with
the Bernoulli potential in the bulk where contributions due to
pairing dominate close to $T_c$.
\end{abstract}
\vskip2pc]
In 1968 Bok and Klein\cite{BK68} have first observed non-zero 
Hall voltage in the Meissner state of a type-I superconductor. 
Instead of Ohmic 
contacts, they have used the Kelvin capacitive coupling between 
the surface and the electrodes. Similar measurements have been 
performed by Brown and Morris\cite{BM68,MB71} and more recently 
by Chiang and Shevchenko\cite{CS86,CS96}. 

Among these experiments, the highest accuracy has been achieved 
by Morris and Brown\cite{MB71}. They found a perfect agreement 
with the prediction of van~Vijfeijken and Staas\cite{VS64} in 
which the original Bernoulli potential (derived by 
Bopp\cite{B37} and made familiar by London\cite{L50}) is extended 
by the 
fountain term of the two-fluid model. In particular, this 
measurement shows no traces of effects predicted later within 
more sophisticated thermodynamical approaches\cite{AW68,R69,H75}.

In this paper we show that the agreement of the two-fluid theory
with experiment is in some sense fortuitous. This theory has 
been developed for the electrostatic potential inside the 
superconductor but it is the surface potential that is measured 
by the Kelvin method. The surface potential is influenced by the 
surface dipole which brings a step-like contribution to the 
electrostatic potential\cite{W96,KW96}. With the surface dipole 
included, the measured surface potential agrees with the 
prediction of the thermodynamic approaches.

The Hall voltage in superconductors is a difference between 
electrostatic potentials, $\varphi$, at the contacts.
All theories agree 
that this
potential has the form of 
the Bernoulli potential,
\begin{equation}
e\varphi\propto -{1\over 2}mv^2,
\label{Bp}
\end{equation}
where $v$ is the velocity of the condensate and $m$ is the mass 
of electrons. The point of disagreement concerns the amplitude of this 
potential. 

London\cite{L50} assumed that the electric force balances the 
inertial and Lorentz forces. The energy-conserving integral of
motion then yields the Bernoulli potential,
\begin{equation}
e\varphi=-{1\over 2}mv^2. 
\label{L1}
\end{equation}
This formula applies only at zero temperature. To cover finite
temperatures, van~Vijfeijken and Staas\cite{VS64} took into account
the fact that the 
normal electrons feel the electric field. Since they stay 
at rest, there must be a balancing force from the 
interaction with the condensate. The reaction force acts on the
condensate and adds to the electric force. This picture leads to
the electrostatic potential which is proportional to the fraction 
of superconducting electrons, 
\begin{equation}
e\varphi=-{n_s\over n}{1\over 2}mv^2.
\label{vS}
\end{equation}

The formula (\ref{vS}) represents the level of theory before the first
successful experiment appeared. With respect to experiments 
it is profitable to express the 
Bernoulli potential (\ref{vS}) as a function of the magnetic field $B$. At 
the surface, $v={e\over m}\lambda B$, where $\lambda=\sqrt{m
\over e^2n_s\mu}$ is the London penetration depth, therefore 
the two-fluid potential (\ref{vS}) reads
\begin{equation}
e\varphi=-{1\over n}{B^2\over 2\mu}.
\label{B1}
\end{equation}
Morris and Brown\cite{MB71} experimentally confirmed the validity 
of (\ref{B1}) in a wide temperature range including the vicinity 
of the critical temperature. 

As discussed by Bok and Klein\cite{BK68}, there is also a simple 
theoretical argument supporting formula (\ref{B1}). In the 
absence of dissipation, the only force acting on the lattice 
is due to the electric field, $-\nabla\varphi$. The total force
per unit area acting on a superconducting slab,
\begin{equation}
F=\int\limits_{{\rm right}}^{{\rm left}} dx\, en\nabla\varphi=
en\varphi_{\rm left}-en\varphi_{\rm right},
\label{B5}
\end{equation}
is proportional to the difference of the surface potentials. 
Here, $-en$ stands for the charge density of the ionic lattice. 
In the magnetic field $B$, the slab with the current density $J$ 
[in A/m] feels the Lorentz force $F=BJ$. Using the Ampere rule, 
$B_{\rm right}-B_{\rm left}=\mu J$, and the mean magnetic field,
 $B={1\over 2}(B_{\rm right}+B_{\rm left})$, the Lorentz force 
can be expressed as the difference between the left and right 
magnetic pressures,
\begin{equation}
F=BJ={B_{\rm right}^2\over 2\mu}-{B_{\rm left}^2\over 2\mu}.
\label{B4}
\end{equation}
One can see that the Lorentz force (\ref{B4}) agrees with 
(\ref{B5}) for potential (\ref{B1}).

In spite of its internal consistency and the satisfactory 
agreement with the experimental data of Bok and Klein, the 
two-fluid theory came under suspicion when 
studies\cite{AW68,R69,H75} based on thermodynamic relations 
and BCS analyses have been accomplished. These approaches result 
in a potential
\begin{equation}
e\varphi=-{\partial n_s\over\partial n}{1\over 2}mv^2.
\label{Ric2}
\end{equation}
For the phenomenological density of condensate,
\begin{equation}
n_s=n\left(1-{T^4\over T_c^4}\right),
\label{Ric3}
\end{equation}
potential (\ref{Ric2}) reads
\begin{equation}
e\varphi=-{n_s\over n}{1\over 2}mv^2+
4{n_n\over n}{\partial\ln T_c\over\partial\ln n}
{1\over 2}mv^2,
\label{Ric4}
\end{equation}
where $n_n=n-n_s$ is the density of normal electrons. The 
first term is identical to the two-fluid potential (\ref{vS}), 
the second is the difference which we will call the thermodynamic 
correction. 

Formulas (\ref{Ric2}) and (\ref{vS}), are very different close 
to $T_c$. Within the BCS relations and the parabolic band 
approximation, one can rearrange 
(\ref{Ric4}) as
\begin{equation}
e\varphi=-{n_s\over n}{1\over 2}mv^2-{4\over 3}
{n_n\over n}\ln{T_D\over T_c}{1\over 2}mv^2.
\label{Ric5}
\end{equation}
Since the Debye temperature $T_D$ is much larger than the critical 
temperature $T_c$, the second term dominates at higher
temperatures, approximately for $T>{2\over 3}T_c$. 

The thermodynamic correction arises from the pairing mechanism.
According to the BCS critical temperature, $T_c\approx T_D
{\rm e}^{-1/{\cal D}V}$, where $\cal D$ is the single-spin density
of states and $V$ is the BCS interaction parameter, the potential 
(\ref{Ric4}) can be written as
\begin{equation}
e\varphi=-{n_s\over n}{1\over 2}mv^2-
{n_n\over n}{1\over {\cal D}V}
{2\over 3}mv^2.
\label{Ric6}
\end{equation}
As speculated by Rickayzen\cite{R69}, the Hall voltage at
temperatures close to $T_c$ can offer an experimental access to 
the BCS interaction $V$. Rickayzen expressed his surprise that 
Bok and Klein\cite{BK68} did not observe any trace of the 
thermodynamic correction at the temperature $T=4.2$~K for lead 
with $T_c=7.2$~K. For $T_D=105$~K the factors in (\ref{Ric5}) 
are comparable, ${n_s\over n}=0.88$ and ${4\over 3}{n_n\over 
n}\ln{T_D\over T_c}=0.42$.

While the measurement of Bok and Klein\cite{BK68} was not
sufficiently precise and the thermodynamic correction could
be lost in the experimental error, later measurement of
Morris and Brown\cite{MB71} clearly proved that the Hall 
voltage is independent of temperature. They also used lead
and explored the temperature range from $4.2$ to $7.0$~K. 
At $T=7$~K the thermodynamic correction dominates the potential,
since the factors in (\ref{Ric5}) are ${n_s\over n}=0.1$
and ${4\over 3}{n_n\over n}\ln{T_D\over T_c}=3.2$. If present,
the thermodynamic correction would amplify the signal thirty
times. This is far beyond the experimental error, estimated to
be less than 10\%. Briefly, the thermodynamic correction
turned out to be absent. It should be mentioned that Morris and 
Brown expected a non-zero thermodynamic correction indicated
by their less precise preliminary measurement\cite{BM68} at 
$T=4.2$~K.

One might feel that the experimental results imply that the 
thermodynamic correction is an ill-defined concept. It is, however, not
the case. The electrostatic potential given by (\ref{Ric2}) 
applies in the bulk. To obtain the potential at the surface, 
one
cannot simply extrapolate this bulk value. At the surface there is a
surface dipole which leads to a discontinuity in the potential.\cite{W96,KW96} 

The surface dipole in a metal introduces a step of several eV's 
in the electrostatic
potential. In the jellium model this step equals to
the Fermi energy $E_F$. With the
exchange and correlation energy included, the surface dipole is
reduced by half which shows the importance of correlations. 
The phase transition into the superconducting state increases 
the correlation energy by the condensation energy. The surface
dipole of the superconductor is thus slightly smaller than the
dipole of the same crystal in the normal state. This difference
is of the order of $\Delta^2/eE_F$.

When a supercurrent flows close to the surface, the condensation 
energy is reduced. The surface dipole thus slightly increases,
reaching the normal state value at the critical current. This
current dependent part of the surface dipole contributes to the
potential observed by the capacitive 
coupling\cite{BK68,BM68,MB71}. Now we show that the surface 
dipole cancels the thermodynamic contribution (second term of
(\ref{Ric4})) to the Hall voltage.

In our formulation we follow Rickayzen\cite{R69}. The increase 
of the free energy due to the current equals the kinetic energy 
of the condensate,
\begin{equation}
\delta{\cal F}=n_s{1\over 2}mv^2.
\label{Ric1}
\end{equation}
The electrochemical potential, $\nu=E_F+\delta\nu_v+e\varphi$, 
is constant in the whole system, therefore $e\varphi=-\delta
\nu_v$. Since $\nu={\partial{\cal F}\over\partial n}$, the 
velocity variation of the local chemical potential is $\delta
\nu_v={\partial\over\partial n}\delta{\cal F}$. Accordingly, 
the electrostatic potential induced by the current is given by
\begin{equation}
e\varphi=-{\partial\over\partial n}\delta{\cal F}.
\label{Ric7}
\end{equation}
With $\delta{\cal F}$ given by (\ref{Ric1}), from (\ref{Ric7}) 
one recovers (\ref{Ric2}).

The step of the electrostatic potential at the surface is 
given by the Budd-Vannimenus theorem\cite{KW96}, 
\begin{equation}
e\varphi_{\rm surf}-e\varphi=n{\partial\over\partial n}
{\delta{\cal F}\over n}.
\label{BVt}
\end{equation}
We note that Budd-Vannimenus theorem has been derived for the
surface potential of the normal metal. Formula (\ref{BVt}) is 
a modification restricted to the potential caused by the 
supercurrent. 

From (\ref{BVt}) and the bulk potential 
(\ref{Ric7}) one finds the electrostatic potential at the 
surface,
\begin{equation}
e\varphi_{\rm surf}=-{n_s\over n}{1\over 2}mv^2.
\label{surfpot}
\end{equation}
This potential is identical to the experimentally confirmed
two-fluid potential (\ref{vS}). Accordingly the surface dipole cancels
exactly the thermodynamic corrections.

In conclusion, the electrostatic potential observed at the
surface of superconductors is determined by the magnetic
pressure. The only material parameter accessible by its
measurement is the charge density of the ionic lattice. In
the bulk, the electrostatic potential depends also on the
pairing mechanism. 

Since the thermodynamic corrections cannot be measured 
on the surface, 
it is desirable to observe the internal electric field directly 
in the bulk of a superconductor. A new experiment in this 
direction has been reported by Kumagai {\em et al}\cite{KNM} 
who have measured the effect of charge density deviations on 
the nuclear quadrupole resonance. 

It is not obvious how the surface potential works in 
layered structures of high-$T_c$ materials. A measurement of
the Hall voltage on these materials, however, involves
enormous technical difficulties.\cite{Tom}

The authors thank Nai Kwong for helpful discussions.
This work was supported by GA\v{C}R 202000643 and 202990410, 
GAAV A1010806 and A1010919 grants. The European ESF program 
VORTEX is also acknowledged.

\end{document}